\input harvmac
\def\p{{\partial}} 
\Title{
\vbox{
\hbox{IFT-P.001/98}  } }
{\vbox{\centerline{\bf 
A Problem with the Superstring Action of Deriglazov and Galajinsky }}} 
\bigskip\centerline{Nathan Berkovits}
\bigskip\centerline{Instituto
de F\'{\i}sica Te\'orica, Univ. Estadual Paulista}
\centerline{Rua Pamplona 145, S\~ao Paulo, SP 01405-900, BRASIL}
\centerline{e-mail: nberkovi@power.ift.unesp.br}
\vskip .3in

Deriglazov and Galajinsky have recently proposed a new covariant
action for the Green-Schwarz superstring which can be constructed
in any spacetime dimension. In this short note, I show 
that their action contains extra on-shell degrees of freedom
as compared with the standard action, and is therefore inequivalent. 

\vskip .2in
\Date{December 1997}

In five recent papers 
\ref\derfi{
A.A. Deriglazov and A.V. Galajinsky, Phys. Rev. D54 (1996) 5195,
hep-th/9604006. }  
\ref\derf{
A.A. Deriglazov and A.V. Galajinsky, Phys. Lett. B381 (1996) 105,
hep-th/9604074.} 
\ref\derth{
A.A. Deriglazov and A.V. Galajinsky, hep-th/9706152.}
\ref\dertwo{
A.A. Deriglazov, hep-th/9709025.} 
\ref\derone {
A.A. Deriglazov and A.V. Galajinsky, to appear in Mod. Phys. Lett. A,
hep-th/9711196.} 
\foot
{The first two papers discuss the problem of
infinitely reducible first class constraints
while the last three papers discuss the $d=11$ superstring.}, 
Deriglazov and Galajinsky have proposed a new
covariant action for the Green-Schwarz superstring which can be
constructed in any spacetime dimension.
Their action is \derone  
\eqn\action{
S = \int d^2 \sigma [ {{- g^{ab}}\over {2\sqrt{-g}}} \Pi^\mu_a \Pi^\mu_b -
i \epsilon^{ab} \p_a x^\mu (\bar\theta \Gamma^\mu \p_b\theta)  }
$$- i\Lambda^\mu \bar\psi \Gamma^\mu \theta - {1\over 
\phi} \Lambda^\mu \Lambda^\mu - \Lambda^\mu \epsilon^{ab} \p_a A^\mu_b],$$
where $\Pi^\mu_a = \p_a x^\mu -i\bar\theta \Gamma^\mu \p_a \theta$. 

The first two terms in \action\ are 
the usual terms in the covariant
Green-Schwarz action \ref\GS{M.B. Green and J.H. Schwarz, Phys. Lett.
B 136 (1984) 367.}. The equations of motion from varying the
fourth and fifth terms in \action\ imply that 
$\Lambda^\mu$ is a constant null vector. Finally, the equation
of motion from varying $\psi$ in the third term eliminates half of
the $\theta$ variables. Since half of the $\theta$ variables are 
eliminated by $\kappa$-symmetry in the standard Green-Schwarz action,
the authors conclude that their action describes the correct
physical degrees of freedom of the superstring without requiring
$\kappa$-symmetry. Furthermore, since their action can be constructed
in any dimension, they seemingly avoid the restriction to $D=3,4,6,10$
which follows from requiring $\kappa$-symemtry. 

However, there is a problem with the action of \action.\foot
{A similar problem was earlier noted by Gates\ref\gat{
S.J. Gates Jr., private communication.}.}  
The problem
is that, although $\Lambda^\mu$ describes a constant
null vector on-shell, the
direction of this null vector cannot be gauge-fixed to point in the
direction $\Lambda^0 = \Lambda^{D-1} =1$, $\Lambda^i =0$ for $i=1$ to $D-2$. 
In other words, in the light-cone gauge $x^0 + x^{D-1} =\tau$, the
restriction on $\theta$ is $\Lambda^\mu \Gamma^\mu \theta =0$, rather
than the usual $(\Gamma^0 - \Gamma^{D-1} )\theta=0$. 

So in addition to 
the usual light-cone degrees of freedom of the Green-Schwarz superstring, 
one also has the $D-1$ global
degrees of freedom 
given by the constant null modes of $\Lambda^\mu$. 
Unlike the functional integration for the standard Green-Schwarz 
superstring, functional integration for the Deriglazov-Galajinsky
superstring 
must include an integration over all
possible choices of these constant null modes. Therefore, there is no
obvious reason why scattering amplitudes using the two different
actions should agree.

\vskip 30pt

{\bf Acknowledgements:} 
I would like to 
thank A. Deriglazov for discussions 
and CNPq grant number 300256/94 for partial financial support.

\listrefs
\end